\documentclass{iau} 
\usepackage{graphicx}
\usepackage{multicol}
\usepackage[labelfont=bf, skip=2pt, font=small]{caption}

\title[On the prospects of imaging Sgr A* from space] 
{On the prospects of imaging Sagittarius A* from space}

\author[Freek Roelofs et al.]   
{Freek Roelofs$^1$, Heino Falcke$^1$$^2$, Christiaan Brinkerink$^1$, Monika Moscibrodzka$^1$, Leonid I. Gurvits$^3$$^4$, Manuel Martin-Neira$^5$, Volodymyr Kudriashov$^1$, Marc Klein-Wolt$^1$, Remo Tilanus$^1$$^6$, Michael Kramer$^2$ \and Luciano Rezzolla$^7$}

\affiliation{$^1$Department of Astrophysics, Institute for Mathematics, Astrophysics and Particle Physics, Radboud University, PO Box 9010, 6500 GL Nijmegen, The Netherlands \\ email: {\tt f.roelofs@astro.ru.nl} \\[\affilskip]
$^2$Max-Planck-Institut f\"{u}r Radioastronomie, Auf dem H\"{u}gel 69, D-53121 Bonn, Germany\\
$^3$Joint Institute for VLBI ERIC, P.O. Box 2, 7990 AA Dwingeloo, The Netherlands\\  
$^4$Department of Astrodynamics and Space Missions, Delft University of Technology, 2629 HS Delft, The Netherlands\\
$^5$European Space Research and Technology Centre (ESTEC), The European Space Agency, Keplerlaan 1, 2201 AZ Noordwijk, The Netherlands\\
$^6$Leiden Observatory, Leiden University, P.O. Box 9513, 2300 RA Leiden, The Netherlands\\
$^7$Institut f\"ur Theoretische Physik, Goethe-Universit\"at, Max-von-Laue-Str. 1, 60438 Frankfurt, Germany
}

\pubyear{2018}
\volume{342}  
\jname{Perseus in Sicily: from black hole to cluster outskirts}
\editors{K. Asada, E. de Gouveia Dal Pino, H. Nagai, R. Nemmen \& M. Giroletti, eds.}
\begin{document}

\maketitle

\begin{abstract}
Very Long Baseline Interferometry (VLBI) at sub-millimeter waves has the potential to image the shadow of the black hole in the Galactic Center, Sagittarius A* (Sgr A*), and thereby test basic predictions of the theory of general relativity. We investigate the imaging prospects of a new Space VLBI mission concept. The setup consists of two satellites in polar or equatorial circular Medium-Earth Orbits with slightly different radii, resulting in a dense spiral-shaped uv-coverage with long baselines, allowing for extremely high-resolution and high-fidelity imaging of radio sources. We simulate observations of a general relativistic magnetohydrodynamics model of Sgr A* for this configuration with noise calculated from model system parameters. After gridding the $uv$-plane and averaging visibilities accumulated over multiple months of integration, images of Sgr A* with a resolution of up to 4 $\mu$as could be reconstructed, allowing for stronger tests of general relativity and accretion models than with ground-based VLBI.
\end{abstract}

\firstsection 
\section{Introduction}
Sagittarius A* (Sgr A*) is a strong radio source in the Galactic Center. Infrared monitoring of stellar orbits in the Galactic Center has led to the conclusion that at the position of this radio source, there is a supermassive black hole with a mass of of $4.3 \pm 0.4 \times 10^6 M_{\odot}$ \cite[(Ghez et al. 2008; Gillessen et al. 2009)]{Ghez2008, Gillessen2009}. Ray-tracing simulations of synchrotron emission originating in an accretion flow led to the prediction of the appearance of a ``black hole shadow'' at millimeter wavelengths \cite[(Falcke et al. 2000)]{Falcke2000}.

The Event Horizon Telescope (EHT) is a mm VLBI array aiming to image a black hole shadow. As the largest black hole in the sky in angular size (53 $\mu$as), Sgr A* is the most promising candidate. The other main target, with a comparable shadow size, is M87. In April 2017, the EHT conducted observations that may lead to the first black hole shadow images. These could inform us about the nature of the accretion flow \cite[(e.g. Dexter et al. 2010; Mo\'scibrodzka et al. 2014; Broderick et al. 2016)]{Dexter2010, Moscibrodzka2014, Broderick2016}, and be used to test general relativity \cite[(e.g. Psaltis et al. 2015; Goddi et al. 2017)]{Psaltis2015, Goddi2017}.

Several observational constraints make setting strong limits on accretion flow parameters and the validity of general relativity a challenging task for the EHT. The array is sparse, making it challenging to produce a robust image. The angular resolution of the EHT is about 23 $\mu$as, set by the observing frequency and the size of the Earth. \cite[Mizuno et al. (2018)]{Mizuno2018} show that this resolution is likely not high enough to distinguish between different theories of gravity, for which a resolution of $\sim5-10$ $\mu$as would be required. At mm and lower wavelengths, the atmosphere introduces strong phase corruptions, making it difficult to push for higher frequencies. For Sgr A*, interstellar scattering distorts the image significantly at 230 GHz. Also, rapid variability poses challenges for imaging this source \cite[(Lu et al. 2016; Johnson et al. 2017; Bouman et al. 2017)]{Lu2016, Johnson2017, Bouman2017}.

\section{Two-satellite Space VLBI concept}
\begin{figure}
\begin{minipage}[t]{.49\textwidth}
    \strut\vspace*{-\baselineskip}\newline\includegraphics[width=\textwidth]{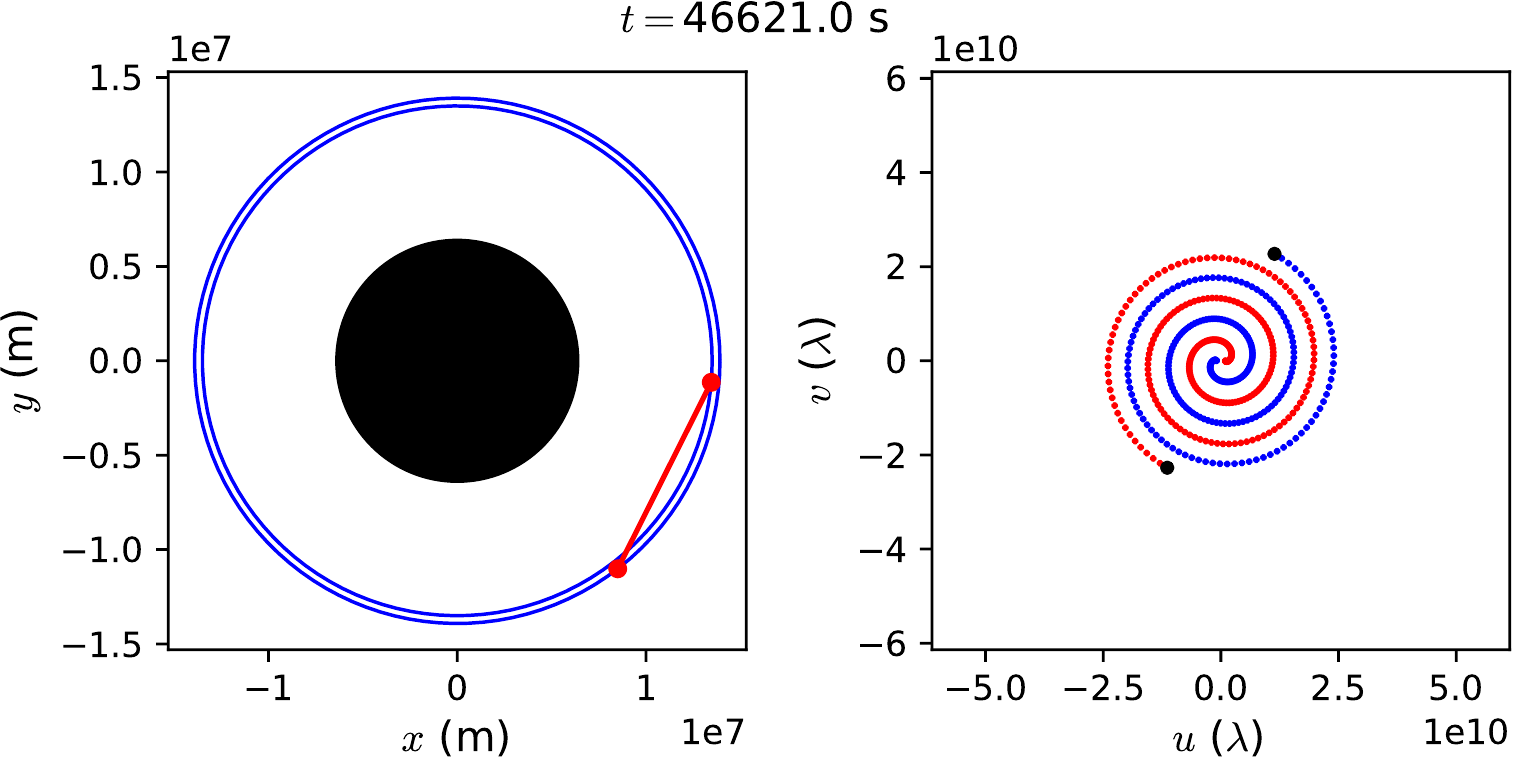}
    \captionof{figure}{Example of satellite positions (left) and $uv$-coverage (right) for circular MEOs. Left, the satellites and baseline are shown in red, the orbits in blue, and the Earth in black. Right, the red and blue points show the $uv$-track for the two baseline directions. The black dots represent current $uv$-coordinates. From the initial satellite positions, the distance between them increases until the Earth occults the ISL. As the inner satellite catches up again, the spiral is traversed inwards.}
    \label{fig:uv}
\end{minipage}
\hfill
\begin{minipage}[t]{.49\textwidth}
    \strut\vspace*{-\baselineskip}\newline\includegraphics[width=\textwidth]{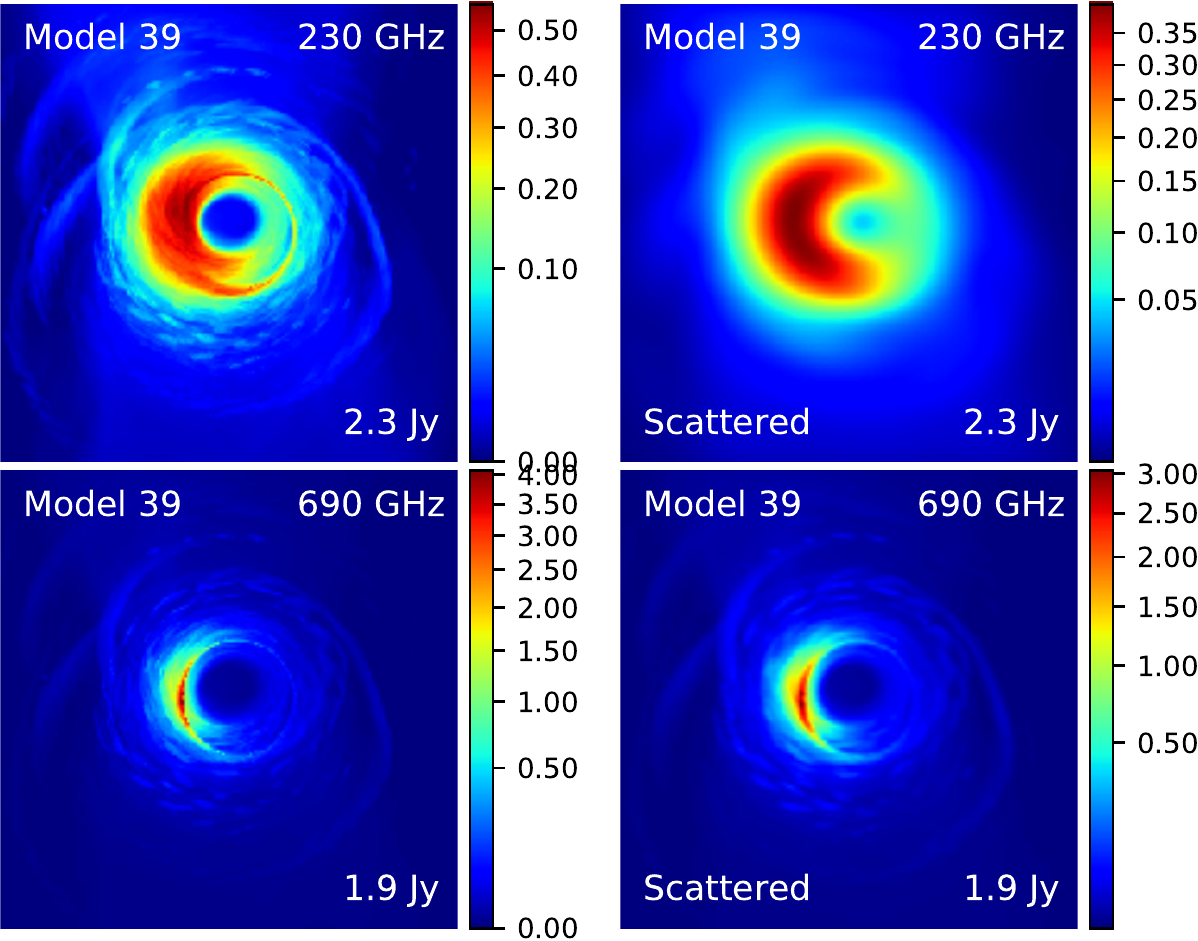}
    \captionof{figure}{Time-averaged GRMHD source models of Sgr A* from \cite[Mo\'scibrodzka et al. (2014)]{Moscibrodzka2014}. Images with note `scattered' were convolved with the scattering kernel from \cite[Bower et al. (2006)]{Bower2006}. The field of view is 210 $\mu$as. Colors indicate brightness/pixel in mJy.}
    \label{fig:models}
\end{minipage}
\end{figure}

The effects mentioned above could be mitigated with a new Space VLBI (SVLBI) concept. An initial design study has been performed for the purpose of the Event Horizon Imager \cite[(EHI, Martin-Neira et al. 2017; Kudriashov et al. 2017)]{Martin2017, Kudriashov2017}. This concept involves two satellites in polar or equatorial Medium Earth Orbits (MEOs) at slightly different radii, observing at frequencies up to $\sim 690$ GHz. This setup results in a dense spiral-shaped $uv$-coverage with long baselines (Fig. \ref{fig:uv}). Because of the absence of the atmosphere, SVLBI allows for observations at higher frequencies, resulting in higher angular resolutions. At high frequencies, the emission originates closer to the black hole event horizon, producing a sharper outline of the black hole shadow due to strong lensing effects (Fig. \ref{fig:models}). Variability of the geometry is expected to be mitigated as the observed image is dominated by lensing effects. Finally, the effect of scattering is mitigated at higher frequencies (Fig. \ref{fig:models}).

Several technical challenges need to be overcome before the concept could turn into an actual mission. In the initial design study of \cite[Martin-Neira et al. (2017)]{Martin2017} and \cite[Kudriashov et al. (2017)]{Kudriashov2017}, correlation takes place on the fly as the satellites share their local oscillator signals over an intersatellite link (ISL). In order to set a delay window that is not prohibitively large, the satellite positions and velocities should be known to within tens of wavelengths. GNSS satellites in combination with ISL ranging measurements may be able to provide sufficient precision, but more research into the feasibility is needed. Also, sufficiently stable clocks should be installed for the interferometer to operate in the submm-regime.

\section{Simulated SVLBI observations}
In order to assess what image quality may be obtained with the EHI, we perform simulated observations of a time-averaged ray-traced general relativistic magnetohydrodynamics (GRMHD) source model of Sgr A* at 690 GHz from \cite[Mo\'scibrodzka et al. (2014)]{Moscibrodzka2014} (Fig. \ref{fig:models}), using the {\tt eht-imaging} library \cite[(Chael et al. 2016)]{Chael2016}. We assume orbital radii of 13,892 and 13,913 km, which gives a maximum baseline length of 5.7$\times 10^{10}\lambda$, corresponding to a 3.6 $\mu$as angular resolution. We blur the source model with the scattering kernel appropriate for Sgr A* \cite[(Bower et al. 2006; Johnson \& Gwinn 2015)]{Bower2006, Johnson2015}. Gaussian thermal noise on the complex visibilities was calculated from model system parameters as $$\sigma=\frac{1}{0.88}\sqrt{\frac{\mathrm{SEFD}_1\mathrm{SEFD}_2}{2\Delta\nu t_{\mathrm{int}}}}, \textrm{ where } \mathrm{SEFD}=\frac{2k_{\mathrm{B}}T_{\mathrm{sys}}}{A_{\mathrm{eff}}},$$ $\Delta\nu$ is the observing bandwidth, $t_{\mathrm{int}}$ is the integration time per measurement, $k_{\mathrm{B}}$ is Boltzmann's constant, $T_{\mathrm{sys}}$ is the system temperature, and $A_{\mathrm{eff}}$ is the effective area of the antenna \cite[(Thompson et al. 2017)]{TMS2017}. The factor 1/0.88 results from 2-bit quantization losses. Here, we adopt $\Delta\nu=2.4$ GHz, $t_{\mathrm{int}} = 180$ s, $T_{\mathrm{sys}}=513$ K, resulting in $\sigma=0.42$ Jy for 3-meter dishes and $\sigma=0.11$ Jy for 6-meter dishes with efficiencies of 0.58.

For our source models, this noise figure allows for a signal-to-noise ratio (S/N) larger than 1 only on the short baselines. In order to get high S/N values on long baselines as well, we apply two strategies. We grid the $uv$-plane, averaging all visibilities in a single cell. Furthermore, we integrate for multiple iterations ($\sim$ months) of the spiral. For these strategies to work in the low-S/N regime, the orbits should be determined precisely enough in post-processing to get coherent phases. Alternatively, the system noise should be reduced significantly so that traditional fringe fitting can be performed. A complication arising with long integrations is the variable character of the source. \cite[Lu et al. (2016)]{Lu2016} showed that for the EHT, an image of the average structure may be obtained by observing multiple days and averaging visibilities at equal $uv$-coordinates. Our strategy of gridding and averaging in the $uv$-plane thus helps mitigating this variability.

\begin{figure}
\begin{minipage}[t]{.49\textwidth}
    \includegraphics[scale=0.225]{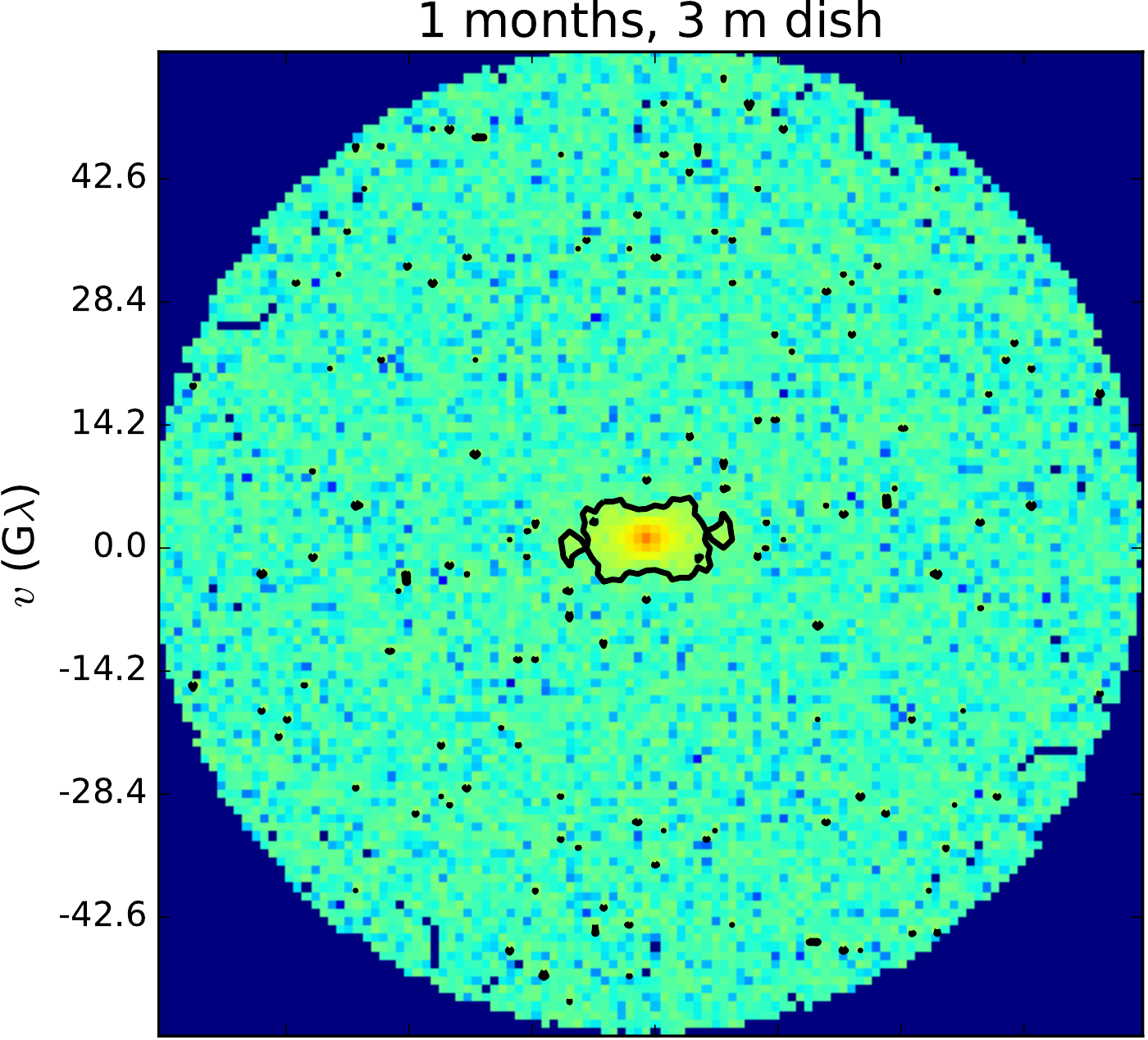}
    \includegraphics[scale=0.225]{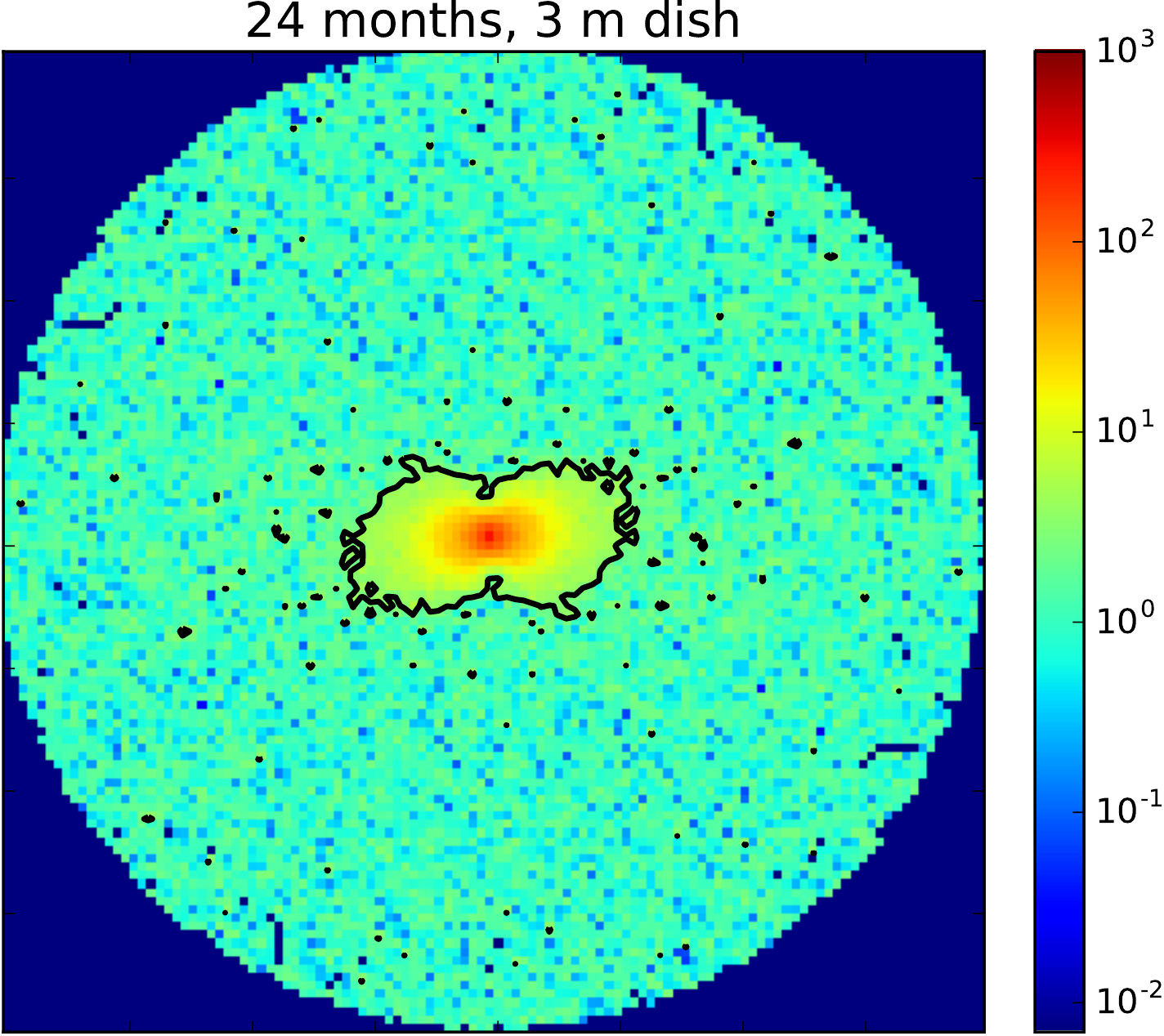}\\
    \includegraphics[scale=0.225]{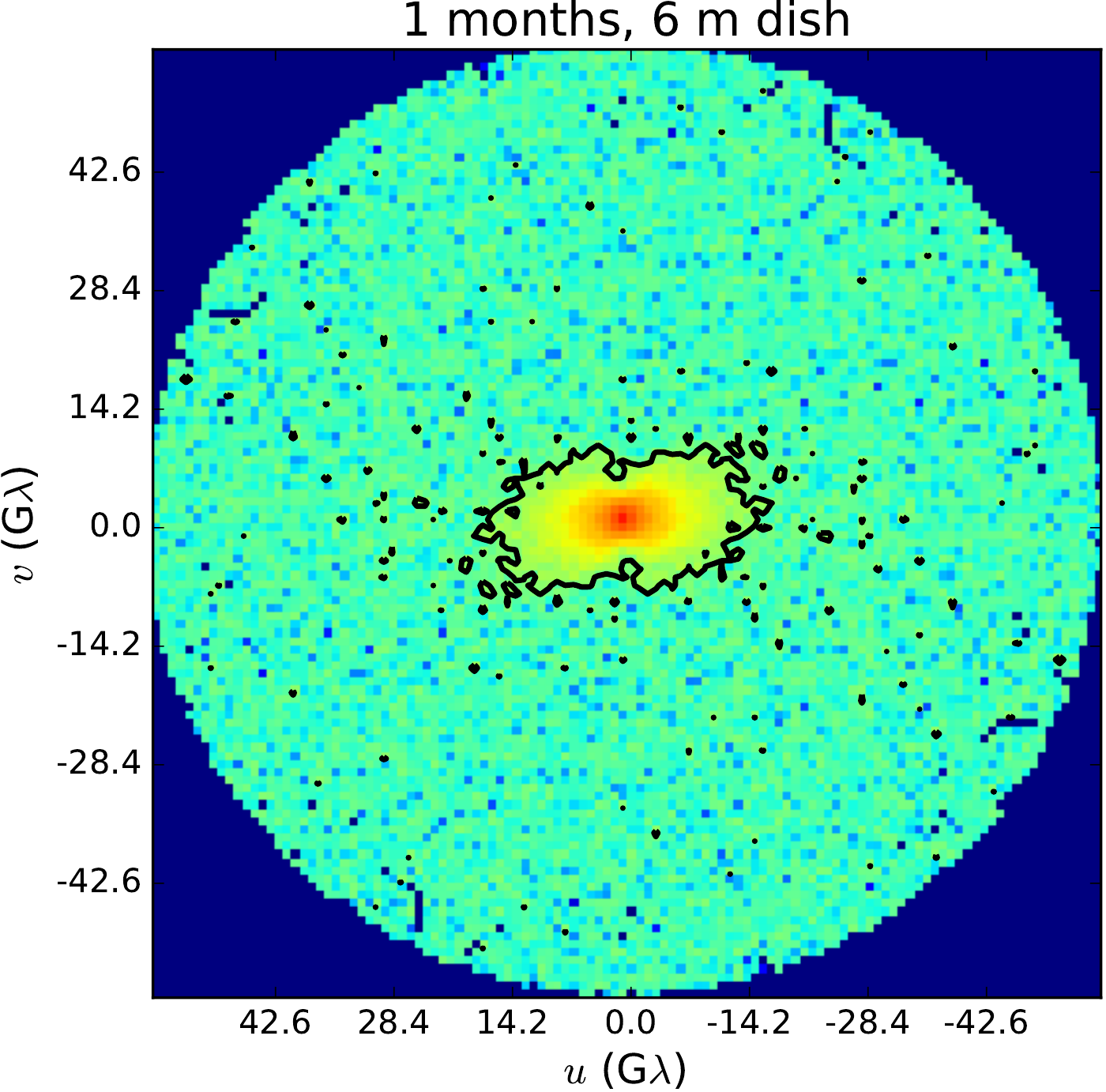}
    \includegraphics[scale=0.225]{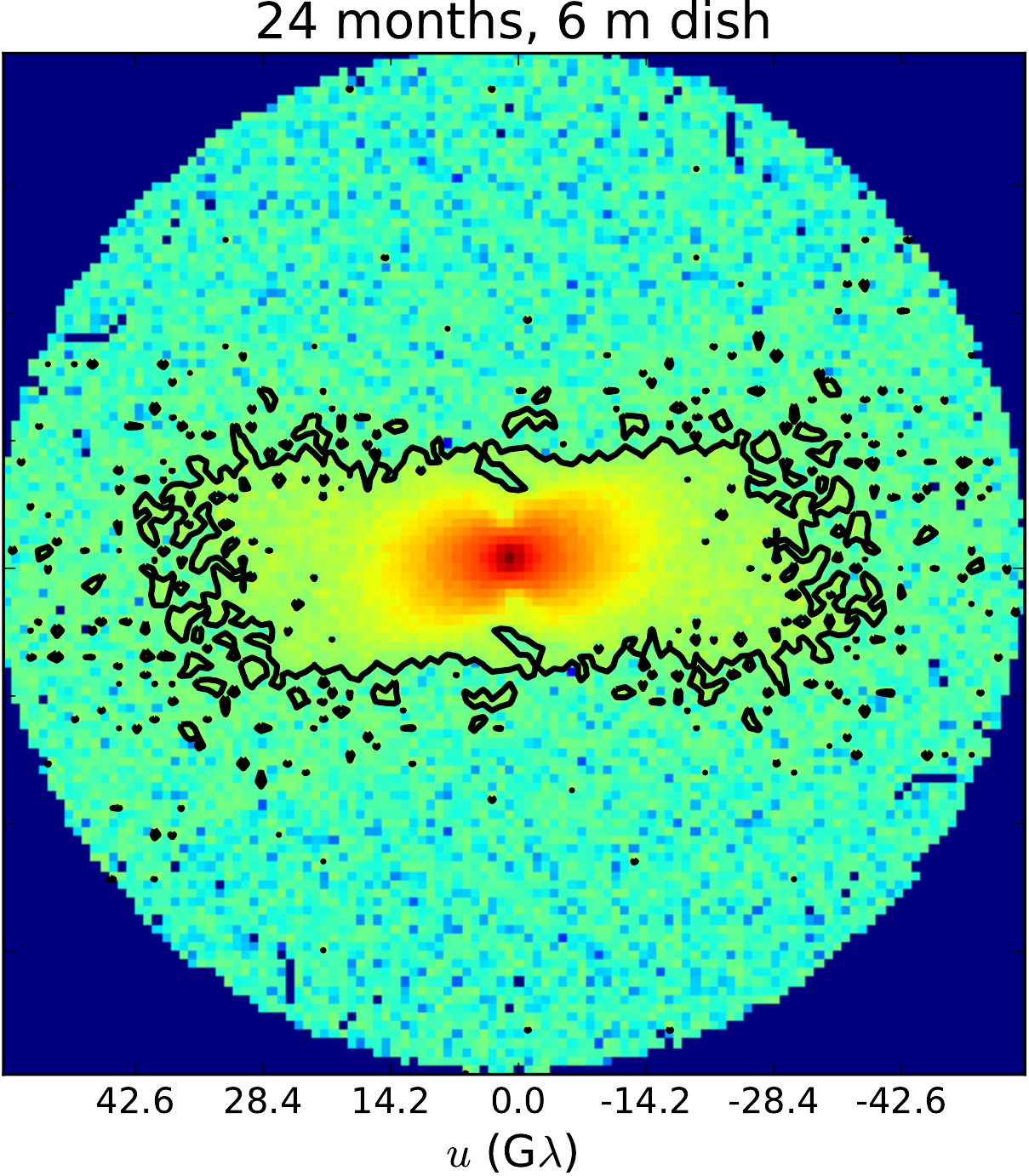}
    \captionof{figure}{S/N map of the gridded visibilities of model 39 (scattered) at 690 GHz after integrating for 1 (left) and 24 (right) months, with a reflector diameter of 3 (upper panels) and 6 (lower panels) meters. Contours indicate the points with an S/N of 3.}
    \label{fig:snr690}
\end{minipage}
\hfill
\begin{minipage}[t]{.49\textwidth}
    \includegraphics[scale=0.225]{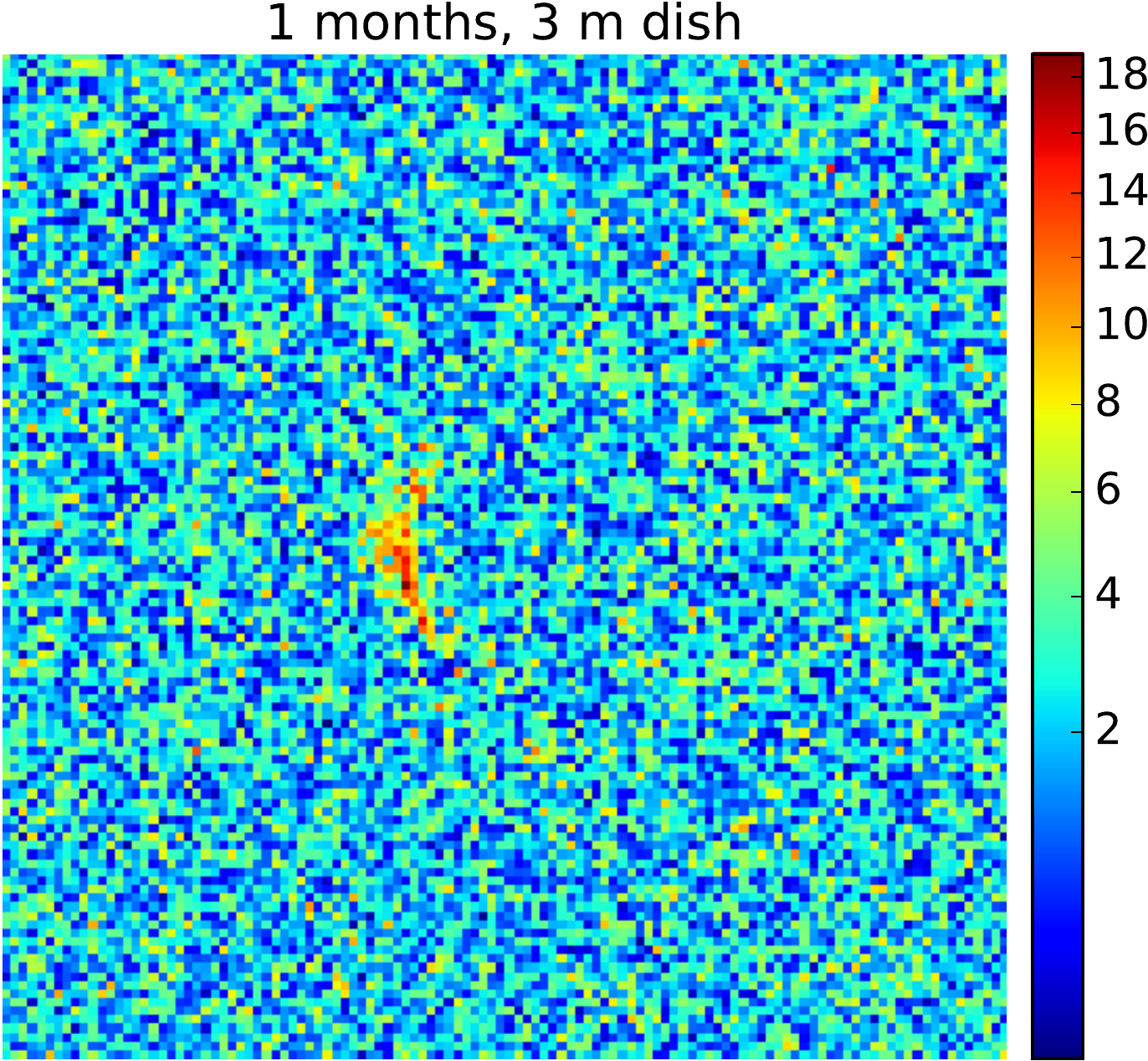}
    \includegraphics[scale=0.225]{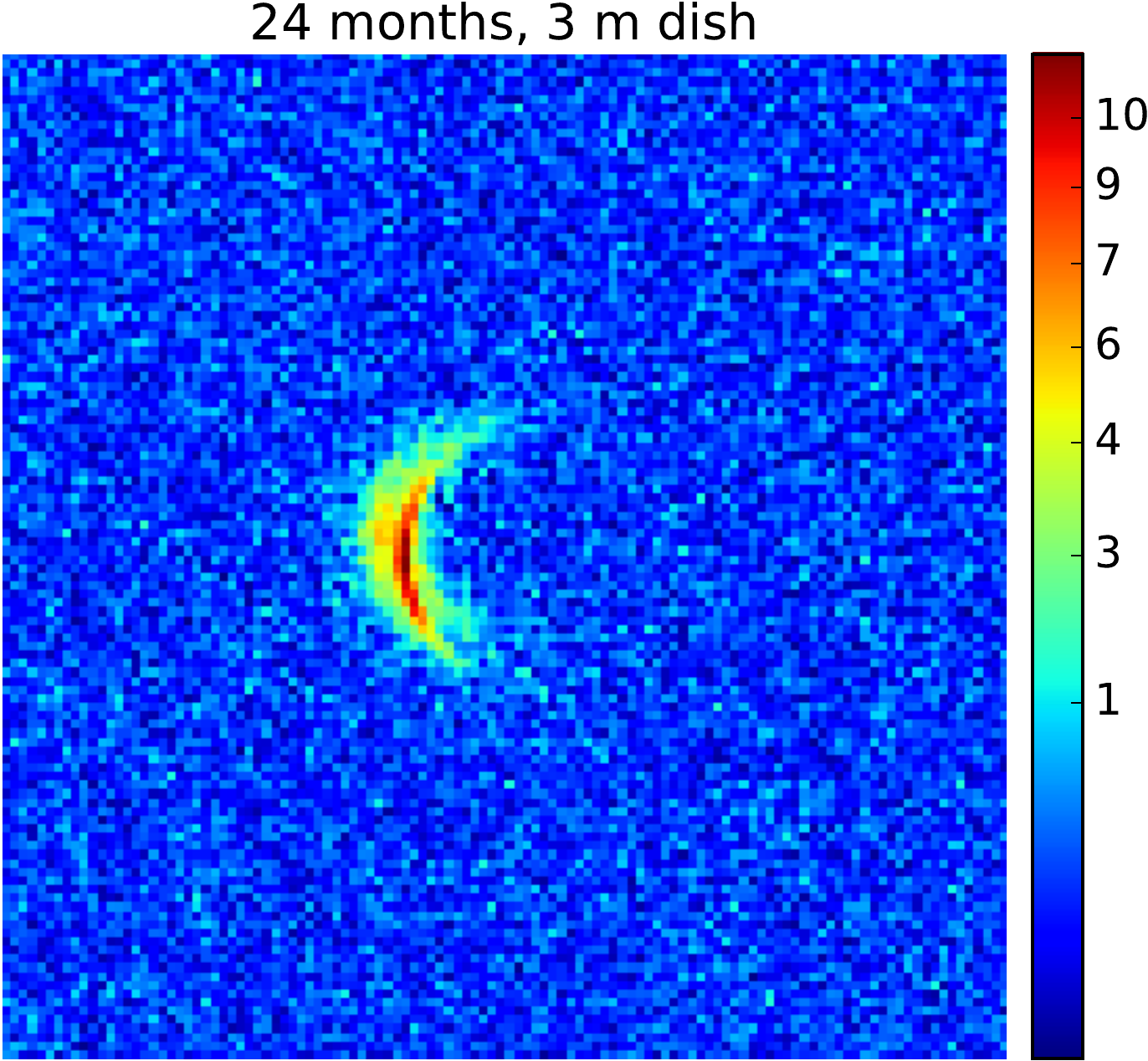}\\
    \includegraphics[scale=0.225]{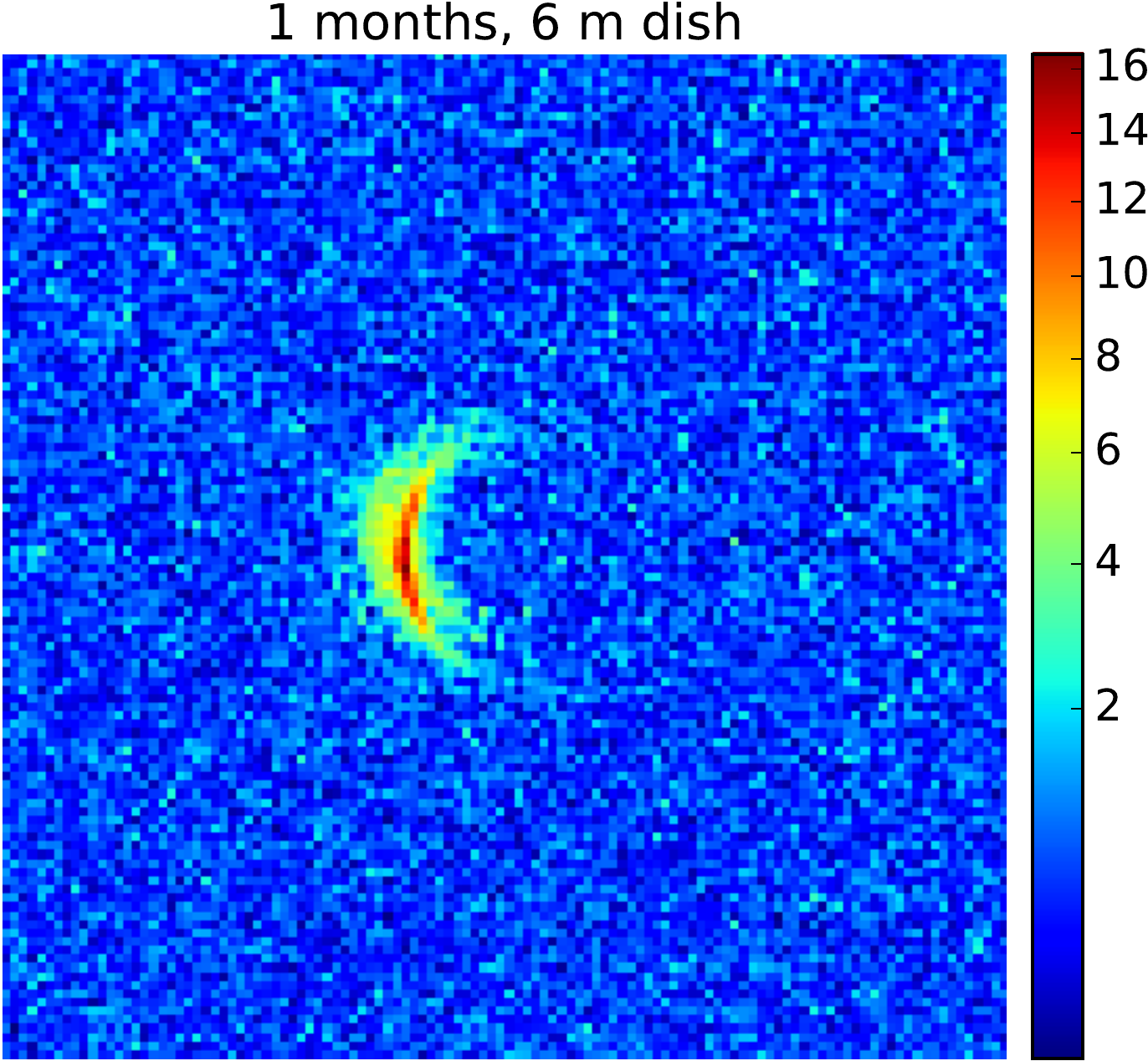}
    \includegraphics[scale=0.225]{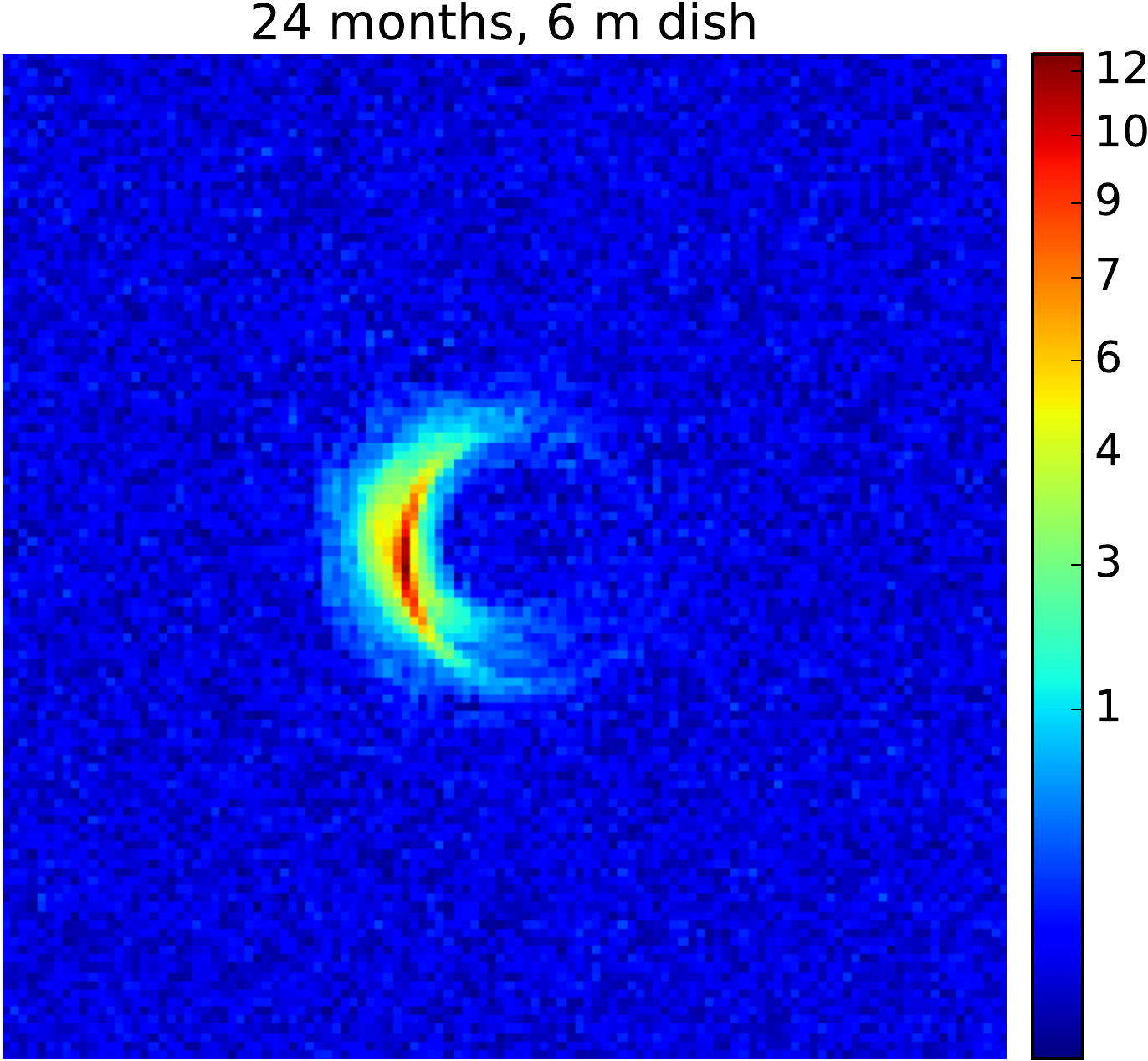}
    \captionof{figure}{FFT of the gridded visibilities of model 39 (scattered) at 690 GHz after integrating for 1, 6, and 24 months (left to right), with a reflector diameter of 3 (upper panels) and 6 (lower panels) meters. The field of view is 210 $\mu$as for all images. Colors indicate brightness/pixel in mJy.}
    \label{fig:fft690}
\end{minipage}
\end{figure}

Figures \ref{fig:snr690} and \ref{fig:fft690} show the resulting S/N $uv$-maps and reconstructed images. Especially with high S/N, the black hole shadow can be reconstructed with an order of magnitude higher resolution than the EHT can achieve. In \cite[Roelofs et al. (2019)]{Roelofs2018}, we show that this is also true for a variable source behind a variable refractive scattering screen, and that the selected orbits allow for sharp images when projection effects due to the source declination of $-29^{\circ}$ are taken into account.

\section{Conclusions and outlook}
We have shown that with the EHI concept, an image of the black hole shadow of Sgr A* could be made with a resolution that is an order of magnitude higher than the EHT, as many limitations of ground-based VLBI could be mitigated. More engineering studies are necessary in order to assess whether the system described here could fly within a reasonable budget. Phase offsets due to uncertainties in the satellite orbits should be included in the simulations when a detailed system model is available. The necessity of using a third satellite to form closure phases, which are immune to these corruptions, should be investigated. Also, the extent to which black hole and accretion parameters can be recovered and general relativity can be tested with this concept as compared to the EHT should be quantified. We also plan to investigate the EHI imaging prospects for other sources such as M87 and other AGN, and polarized GRMHD models \cite[(e.g. Gold et al. 2017; Mo\'scibrodzka et al. 2017)]{Gold2017, Moscibrodzka2018}, using different orbital configurations for filling the $uv$-plane on different timescales.\\

This work is supported by the ERC Synergy Grant ``BlackHoleCam'' (Grant 610058).

\newcommand{\apj}{ApJ}
\newcommand{\apjl}{ApJL}
\newcommand{\aap}{A\&A}
\newcommand{\mnras}{MNRAS}

\end{document}